\documentclass[a4paper]{jpconf}
\usepackage{graphicx}
\begin{document}
\title{Seeding self-modulation of a long proton bunch with a short electron bunch}

\author{P. Muggli,$^1$ P.~I. Morales Guzman$,^{1,3}$ A.-M. Bachmann,$^{1,2,3}$ M. H{\"u}ther,$^1$ M. Moreira, $^{2,4}$ M. Turner,$^2$ and J. Vieira$^4$}

\address{$^1$Max Planck Institute for Physics, %F{\"o}hringer Ring 6, 
Munich 80805, Germany}
\address{$^2$CERN, Geneva 1211, Switzerland}
\address{$^3$Technical University Munich, %James-Franck-Str. 1,
Garching 85748, Germany}
\address{$^4$IST, Lisbon 1049-001, Portugal}

\ead{muggli@mpp.mpg.de}

\begin{abstract}
We briefly compare in numerical simulations the relativistic ionization front and electron bunch seeding of the self-modulation of a relativistic proton bunch in plasma. %
When parameters are such that initial wakefields %at the peak of the proton bunch 
are equal with the two seeding methods, the evolution of the maximum longitudinal wakefields along the plasma are similar. % 
We also propose a possible seeding/injection scheme using a single plasma that we will study in upcoming simulations work. %
\end{abstract}

\section{Introduction}

Seeding a process consists in providing a signal level larger than the noise level present in the system, from which the process could uncontrollably grow. %
With seeding, the process grows from a known and controllable initial signal. %
The final state of the system may be the result of non-linear evolution, but when the process is seeded, the final state is uniquely related to the seed signal both in amplitude and phase. %
In the case of self-modulation (SM) of a relativistic (v$_b\cong$ c), charged particle bunch in plasma~\cite{bib:kumar}, the process is the transformation of the long, continuous bunch into a train of shorter micro-bunches through the effect of the periodic transverse wakefields. %
The important wakefields final parameters are their amplitude and their relative phase with respect to the seed point. %
This is key for external injection of a bunch of particles to be accelerated by these wakefields. %
This bunch must be short with respect to the wakefields period. %
It must be injected in a narrow phase range of accelerating and focusing wakefields, whose extent is on the order of one quarter wakefields period (plasma wakefields linear theory~\cite{bib:keinings}). % 
%Reproducibility of the proton bunch modulation amplitude means reproducibility of the electron bunch final energy. %
Reproducibility of the wakefields amplitude and phase means that the electron bunch can be injected with a fixed time delay (or phase) with respect to the seed signal. %
Reproducibility is necessary for loading of the wakefields and in general to be able to produce a quality accelerated bunch, eventually using feedback systems, as in conventional accelerators. %

The period of (linear) wakefields in an initially neutral plasma of uniform electron density n$_{e0}$ is given by the plasma wavelength $\lambda_{pe}=2\pi v_b/\omega_{pe}\cong 2\pi c/\omega_{pe}$. %
Here $\omega_{pe}=\left(\frac{n_{e0}e^2}{\epsilon_0m_e}\right)^{1/2}$ is the angular plasma electron frequency.\footnote{Here $e$ and $m_e$ are the electron charge and mass, and $\epsilon_0$ is the vacuum permittivity.} % %~\cite{bib:omegape}. %
The bunch is considered long, when longer than $\lambda_{pe}$: $\sigma_z\gg\lambda_{pe}$, $\sigma_z$ the root-mean-square (RMS) length of a Gaussian bunch. %
Micro-bunches are separated by, and are shorter than $\lambda_{pe}$. %
For the bunch to be stable against the current filamentation instability~\cite{bib:CFI} and to effectively drive wakefields, its transverse size must also be smaller than the cold plasma collisionless skin depth $c$/$\omega_{pe}$. %
In the following, size comparisons refer to $\lambda_{pe}$ or $c$/$\omega_{pe}$. % 

\section{Self-Modulation Seeding}

Methods to provide seed wakefields include: short, intense laser pulse or particle bunch preceding the bunch to self-modulate; sharp rise of the long bunch density; and relativistic ionization front traveling with and within the long bunch. %~\cite{bib:seeding}????
Each method has pros and cons. %
In experiments, the seeding method is chosen with respect to practical considerations, e.g. availability of a laser pulse or particle bunch, long bunch parameters, plasma creation process, etc. %

Producing a sharp rising edge in the long bunch is practical with low energy particles~\cite{bib:mask}. %
Experiments have shown that a 60\,MeV electron bunch drives multiple wakefields periods along itself~\cite{bib:fang}. %
These can serve as SM seed wakefields. %
However, for bunches of high-energy particles (e.g., 400\,GeV protons), this seeding method requires parameters and means (relative energy spread at the \%-level, magnetic dog-leg or chicane) that are proportional to the particles' inertia ($\gamma$m, $\gamma$ their relativistic factor) and quickly become prohibitive in size and cost. %~\cite{bib:guoxing}??

\subsection{Relativistic Ionization Front Seeding}

The relativistic ionization front method %~\cite{bib:dan}? 
requires a short laser pulse co-propagating within the proton bunch, and a low ionization potential gas or vapor to keep the laser pulse intensity relatively low. %
In this case, it is the fast creation of plasma and the onset of beam plasma interaction within the bunch that drives the seed wakefields. %  
This method was very successfully used in the AWAKE experiment%~\cite{bib:awake}
~\cite{bib:muggli} with a rubidium vapor of density (1-10)$\times$10$^{14}$\,cm$^{-3}$ and with a laser pulse 120\,fs-long  ($\lambda_0 = $ 780\,nm) and an intensity I$_0\sim$10\,TW/cm$^2$ to fully ionize the atoms of their first electron~\cite{bib:karl,bib:marlene}. %
We note that with these parameters the value of the laser pulse a$_0\cong8.6\times10^{-10}\lambda_0[\mathrm{\mu m}]I_0^{1/2}$\,[W/cm$^2$] reaches only $\cong$ 0.01. %
This places laser wakefields excitation by the laser pulse in the linear regime. %
That means that the longitudinal wakefields amplitude the laser pulse drives is on the order of $\cong\frac{\pi}{4}\frac{a_0^2}{2}E_{WB}$~\cite{bib:esarey}, i.e., 100\,kV/m, where E$_{WB}=m_ec\omega_{pe}/e$ is the cold plasma wave breaking field (n$_{e0}=7\times10^{14}\,\mathrm{cm}^{-3}$). %
This amplitude is much smaller than that of the wakefields at the ionization front, which is on the order of MV/m (see below). %
Thus SM seeding occurs because of the sudden onset of beam-plasma interaction at the ionization front, not because of the wakefields driven by the laser pulse. %
%For example, experiments show no evidence of seeding when placing the laser pulse ahead of the proton bunch~\cite{bib:fabian}. %

In the relativistic ionization front seeding method, the seed wakefields amplitude can be defined as the wakefields driven by the bunch at the location of the ionizing laser pulse or ionization front. %
In the case of a long bunch, with low density n$_b\ll n_{e0}$, the seed wakefields amplitude can be calculated from linear plasma wakefields theory~\cite{bib:keinings}, considering the bunch density does not change over a few wakefields periods behind the ionization front. %
In this case, the seed amplitude is $W_{\perp}=2\frac{en_{b\xi_s}}{\epsilon_0k_{pe}^2}\frac{dR}{dr}|_{r=\sigma_r}$, where $R(r)$ reflects the transverse dependency of the wakefields on the bunch transverse profile. %ref or expression?
The bunch density at seed location $\xi_s$ is $n_{b\xi_s}$. %
When $k_{pe}\sigma_r\cong1$, $k_{pe}=\omega_{pe}$/c, $\frac{dR}{dr}|_{r=\sigma_r}\cong k_{pe}$. % 
For parameters of AWAKE, the seed wakefields amplitude reaches a few MV/m~\cite{bib:marlene} and exceeds that driven by the laser pulse when placed near the bunch peak density location. % numbers?
The longitudinal wakefields amplitude is $W_{\parallel}=\frac{en_{b\xi_s}}{\epsilon_0k_{pe}}R(r)$ and reaches values similar to those of $W_{\perp}$. %

This seeding method has a number of favorable characteristics. %
First, it only requires a laser pulse short when compared to the wakefields period and intense enough to ionize a gas or vapor. %
This is easily satisfied with lasers available today when operating at low plasma density and with a low ionization potential vapor, e.g., alkali metals~\cite{bib:oz}. %~\cite{bib:josh}?. %
We note here that ionization occurs on a time scale on the order of the laser field period, even shorter than the pulse length in most cases. %~\cite{bib:keldysh}?. %   
Second, the seed wakefields are driven by the bunch to self-modulate. %
That means that there is natural alignment between the seed wakefields and the bunch. %
This is important so as not to seed the non-axisymmetric hose instability (HI) mode~\cite{bib:witthum,bib:mathias}. %
Third, the seed wakefields naturally have the same transverse structure as that driven by the bunch to self-modulate. %
This may not be the case when seed wakefields are generated by a driver preceding the long bunch. %
Fourth, a replica of the ionizing laser pulse, thus (perfectly) synchronized with it, can be used to drive an RF gun that provides a synchronised electron bunch for external injection. %

The laser pulse serves two purposes that can be decoupled: plasma creation by itself and, in conjunction with the drive bunch, wakefields seeding. % 
Ionization depletes the energy of the laser pulse and thus limits the plasma density/length product that can be created by this ionization method. %
This directly limits the maximum energy gain by witness bunch particles externally injected into the wakefields. %

To avoid staging and its intricacies, acceleration in a single, preformed and very long plasma may be desirable. %~\cite{bib:staging}??
Also, as the wakefields phase velocity varies during growth of the SM process~\cite{bib:pukhov}, external injection of electrons to be accelerated at a location downstream from the SM saturation point may be desirable. %
This naturally calls for the use of two plasma sections: one for self-modulation of the proton bunch, and one for acceleration of the electron bunch, sections separated by an injection region. %

The ionization front seeding method leaves un-modulated the fraction of the long bunch ahead of the seed point, since it travels in neutral vapor or gas. %
However, propagation of the bunch with its front un-modulated in a following preformed plasma means that self-modulation instability may grow in the bunch front. %
This growth would generate wakefields that would add to the ones driven by the self-modulated back part of the bunch. %
Since the SMI in the bunch front is not seeded, the phase of those wakefields is not controlled and may be different from event to event. %
By causality, these wakefields would also perturb the self-modulated bunch train itself and probably prevent acceleration and quality preservation of the injected electron bunch. %
Self-modulation instability of the whole bunch in a preformed plasma, in this case with the ionizing laser pulse placed ahead of the proton bunch, was observed experimentally~\cite{bib:spencer}. %
A seeding method that leads to self-modulation of the entire bunch may therefore be required. %

\subsection{Electron Bunch Seeding}

We consider here self-modulation seeding using an electron bunch. %
This is natural since an electron accelerator is necessary for external injection. %
Also, driving large amplitude wakefields over a few meters of plasma with a laser pulse may be challenging. %
For effective seeding, the electron bunch may have to drive wakefields with the same transverse structure as that driven by the long bunch. %
Therefore, the electron bunch may have to have the same transverse size as the un-modulated, long bunch. %
This places constraints on the bunch charge since the transverse size of the wakefields and thus of the drive bunch may be quite large at low plasma density, i.e., 200\,$\mathrm{\mu}$m at n$_{e0}=7\times$10$^{14}$\,cm$^{-3}$. %
Misalignment between the electron and the long bunches may act as seed for the HI of the long bunch. %
This could be somewhat mitigated by making the electron bunch radius larger than that of the long bunch, though global misalignment would persist. %
Also, since the long bunch satisfies k$_{pe}\sigma_{rp^+}\cong$1, theory predicts that the electron bunch with k$_{pe}\sigma_{r}>$ 1 is subject to the current filamentation instability (CFI)~\cite{bib:CFI}. %
Experiments with a low energy electron bunch showed that CFI only occurred for k$_{pe}\sigma_{r}>$ 2.2, i.e., when at least two filaments can be formed out of the original bunch~\cite{bib:CFIexp}. %
%A larger electron bunch radius may still relax the alignment tolerance or sensitivity of the proton bunch to HI development while seeding SM. %

The length of the seed bunch must be $<\lambda_{pe}/2$ and its charge sufficient to drive the required seeding amplitude value. %
The long bunch has an initial density n$_{b0}$ smaller than that of the plasma (over-dense plasma). %
Seed wakefields, for example when seeding with an ionization front, are thus in the linear plasma wakefields theory regime (n$_{b0}\ll n_{e0}$~\cite{bib:keinings}). %
When using an electron bunch to seed the wakefields, electron bunch parameters, such as density and length, can be chosen independently from those of the proton bunch. %
For example, they can be chosen so the bunch drives non-linear wakefields, possibly in the blow-out regime. %
The development of seeded self-modulation (SSM) from wakefields with amplitude equal or larger than that driven by the self-modulated wakefields may be quite different than from linear wakefields. %
In particular, the saturation length may be shorter and issues related to wakefields phase velocity during the growth of the SSM mitigated~\cite{bib:pukhov}. % 

Also, the linear transverse seed wakefields are focusing (or null) over most of the bunch length. %
These develop into focusing and defocusing wakefields during the growth. %
Transverse  seed wakefields driven by the electron bunch, i.e., behind the electron bunch placed ahead of the proton bunch are already periodically focusing and defocusing. %
This affects the development of the SM process. %

A major issue may also be that, depending on the electron bunch parameters, the seed wakefields can be present over a distance much longer than the growth or saturation length of the SM process. %
Though of lower amplitude than that of the fully developed self-modulated wakefields, these can also interfere with acceleration over long distance. %
For example, their phase remains constant over the plasma length whereas that of the self-modulated bunch drifts backwards over the growth region~\cite{bib:pukhov}. %
One solution may be to choose the seed bunch energy so that it is depleted by the seed wakefields (its own) over a length on the order of the SM-process saturation length or on the order of the length necessary for the self-modulating bunch to drive wakefields of amplitude comparable to that of the seed ones. %

Transverse evolution of the seed bunch and possible matching to the plasma focusing force are other topics of research. %

\section{Seeding Methods Comparison}

\begin{figure}[ht]
\centering
\includegraphics[scale=0.5]{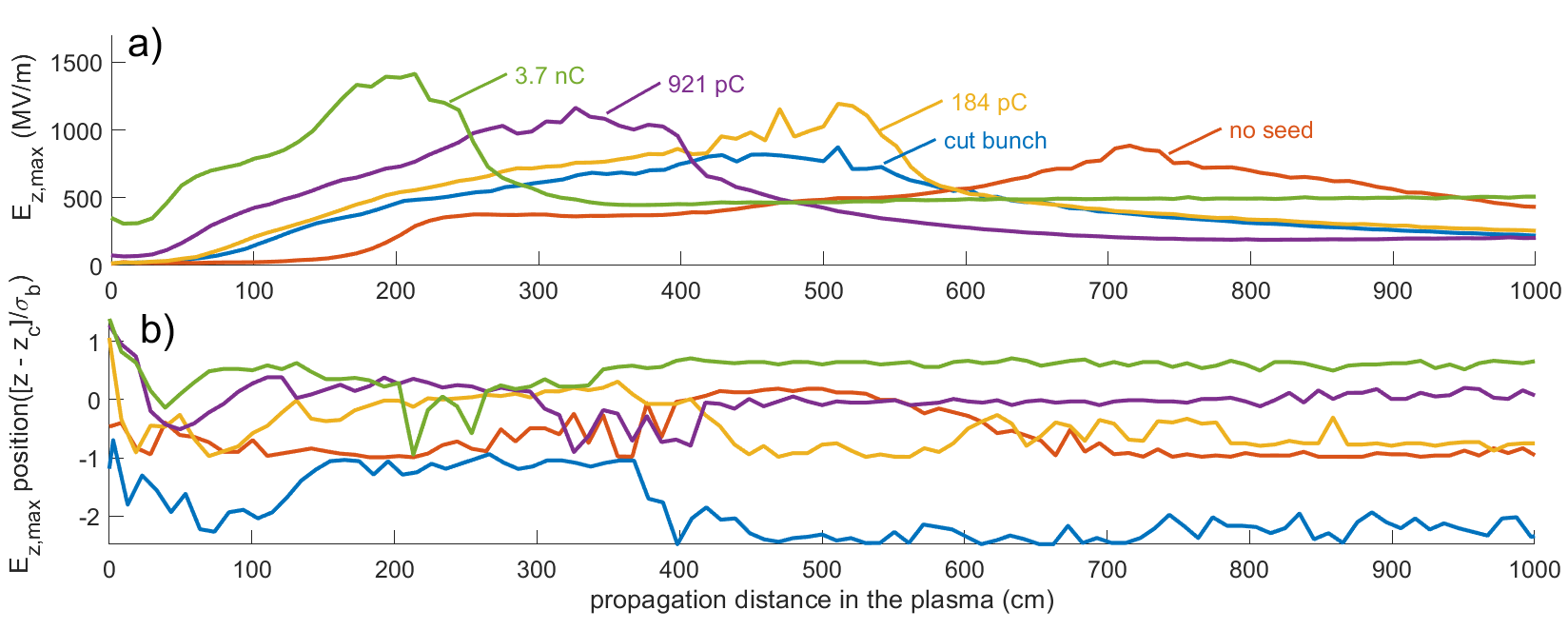}
\caption{a) Maximum longitudinal electric field E$_{z,max}$ as a function of position along the plasma for seeding: bunch cut placed in the center of the proton bunch (blue line); bunch cut at three RMS lengths ahead of the bunch center (orange line, labelled as ``no seed"); short particle bunch placed ahead of the proton bunch, Q = 184\,pC, yellow line, $Q$ = 921\,pC, purple line, and $Q$ = 3.69\,nC, green line. 
%Seed wakefields amplitude is shown in parentheses in the legends.
b) Position of E$_{z,max}$, with respect to the bunch center ($z_\mathrm{c}$), in terms of the bunch RMS length ($\sigma_\mathrm{b}$). Same color code for the lines.
Since all these simulations were performed with the same plasma density and beam sizes, the same numerical parameters are used: $\Delta_r= 0.01\,c/\omega_{pe}$,  $\Delta_z=0.015\,c/\omega_{pe}$,  $\Delta_t= 0.0074\,/\omega_{pe}$ and $4 \times 4$ particles per cell for the plasma electrons. %
The simulation box size is  $7.97\,c/\omega_{pe}$ in radius and $1204.89\,c/\omega_{pe}$ in length for all cases except the cut bunch case, in which it is $746.84\,c/\omega_{pe}$. %
} 
\label{fig:Ezmaxofz} 
\end{figure}

Using particle-in-cell simulations with OSIRIS~\cite{bib:osiris} in 2D cylindrical coordinates, we briefly compare seeding of the SM process of a proton bunch with a sharp cut, representing seeding by an ionization front, and by a preceding particle bunch. %
We choose the cut bunch case as reference. %
The proton bunch has a Gaussian longitudinal profile with a root-mean-square (RMS) length of $\sigma_b=$ 6\,cm (200\,ps duration) and a bi-Gaussian transverse profile with a RMS radius of 200\,$\mathrm{\mu}$m. %
Its charge is 48.1\,nC, and it has a relativistic factor of $\gamma_{p^+} = 426$. %
Its normalized emittance is 3.6\,mm-mrad.  %
For this case we place the cut in the center of the proton bunch (the point where its density is highest). %
The plasma density is 7$\times$10$^{14}$\,cm$^{-3}$. %
This yields a seed amplitude that we define as the longitudinal wakefields amplitude immediately behind the cut of 13.5\,MV/m (or 5$\times$10$^{-3}\,$E$_{WB}$, E$_{WB}$=2.55\,GV/m). % I THINK WE SHOULD USE THE TRANSVERSE WAKEFIELDS AMPLITUDE
We note here that we quote longitudinal wakefields amplitudes because they can always be evaluated on the system axis. %
Transverse wakefields amplitudes are zero on axis, and depend on the radius where they are evaluated (usually at $r=\sigma_r$). %
However, since the bunch and the micro-bunches radii change along the SM process, quoting such amplitudes becomes ambiguous. %
The maximum longitudinal electric field along the bunch as a function of propagation in the plasma, together with the position along the bunch where that amplitude is reached, are plotted on Fig.~\ref{fig:Ezmaxofz} as the blue lines. % 
The maximum wakefields amplitudes grows from its seed value, saturates around $\mathrm{z}=450$\,cm at $\sim$820\,MV/m (0.32\,E$_{WB}$) and decays after that. %HERE AGAIN, TRANSVERSE FIELDS AMPLITUDE
The position of this maximum amplitude shifts quickly from the bunch front to approximately two RMS lengths behind that point for the first meter and a half, after which it remains close to one RMS length behind the seeding position, until the maximum field starts decreasing. %
In the decreasing portion, the maximum is found around two and a half RMS lengths from the seeding position. %

We now consider a seed electron bunch that drives wakefields with a constant amplitude, which is equal to that of the cut bunch. %
We note that, in this case, the sum of the seed bunch wakefields and the long bunch adiabatic response are equal to the wakefields of the cut bunch at the peak of the bunch current profile (the cut bunch seed point). % 
Also, the seed bunch drives defocusing fields along the proton bunch ahead of the bunch center, a major difference from the ionization front seeding case. %
It is clear that the two situations are not identical, however, we consider them as sufficiently similar to draw some important conclusions. %
The maximum electric field and its position along the bunch for this seeding case are also plotted on Fig.~\ref{fig:Ezmaxofz} as the yellow lines. %
The seed bunch we consider has a negative charge, and we increase the mass of its particles by ten orders of magnitude compared to the mass of an electron. %
This is a way to avoid transverse evolution and dephasing of the seed bunch along the simulation that would occur with a low energy electron bunch not matched to the plasma focusing force. %
The RMS length is chosen to be 300\,$\mathrm{\mu}$m, short when compared to the wakefields period (1270\,$\mathrm{\mu}$m). %
The bunch has a Gaussian longitudinal profile and bi-Gaussian transverse profile with a size of 200\,$\mathrm{\mu}$m, equal to that of the proton bunch. %
The energy of the seed bunch particles is sufficient to drive a seed wakefields amplitude over the whole propagation distance (10\,m) without dephasing with respect to the wakefields. %
%The bunch emittance is chosen low enough to avoid strong divergence over a distance in vacuum comparable to the seed distance. %
The seed bunch is placed 631\,ps (or about three proton bunch RMS lengths) ahead of the proton bunch peak density point. % 

In this case, the wakefields also grow and saturate. %
However, they reach a higher peak value at saturation, $\sim$1200\,MV/m (0.47\,E$_{WB}$). % rather than $\sim$820\,MV/m. %
Saturation is reached around $\sim$500\,cm. % rather than $\sim$4.5\,m. %
The fields also decrease after saturation. %
The position of this maximum field follows the same trend as that with cut bunch seeding, but around one RMS length closer to the bunch center. %
The evolution of the maximum field is not significantly different between the two cases, especially considering plasma lengths necessary for saturation of fields and distances useful for acceleration, i.e., after saturation. %

We now consider the case of seeding with five times the charge of the previous case. %
As a result of the charge increase, the seed wakefields amplitude is $\sim$65.2\,MV/m (0.025\,E$_{WB}$), with results plotted as the purple lines on Fig.~\ref{fig:Ezmaxofz}. %
The maximum value is similar to that of the previous seeding case, $\sim$1100\,MV/m (0.43\,E$_{WB}$), and is reached a bit earlier, at z = 330\,cm. %
However, the field decreases in the same manner as in the previous two cases, and reaches about the same amplitude at 10\,m. %
%The location along the bunch of the maximum field remains close to the bunch center, changing to about half an RMS length behind it for the zone around the peak of the maximum longitudinal fields. % 

We further use a seed bunch with a charge increased by twenty times compared to the first seed bunch charge, with results plotted on Fig. \ref{fig:Ezmaxofz} as the green lines. %
In this case, the seed wakefields amplitude is 298\,MV/m (0.12\,E$_{WB}$). %
Since the seed bunch charge density is approximately 17\% of the plasma density, seed wakefields are in the quasi-linear regime, giving an amplitude 19\% higher than that predicted by linear theory. %
The maximum longitudinal field is reached even earlier than in the previous cases, at about 200\,cm, and approaches $\sim$1400 MV/m (0.5\,E$_{WB}$), also higher than before. %
After the peak, the train of micro-bunches reaches a stable configuration with a maximum wakefields amplitude of $\sim$495\,MV/m (0.19\,E$_{WB}$), only about 200\,MV/m higher than the seed level. %
The position of the maximum field remains at half a RMS length in front of the bunch center, starting at about 400\,cm. %, ideal to inject an electron bunch for acceleration. %

\begin{figure}[ht]
\centering
\includegraphics[scale=0.25]{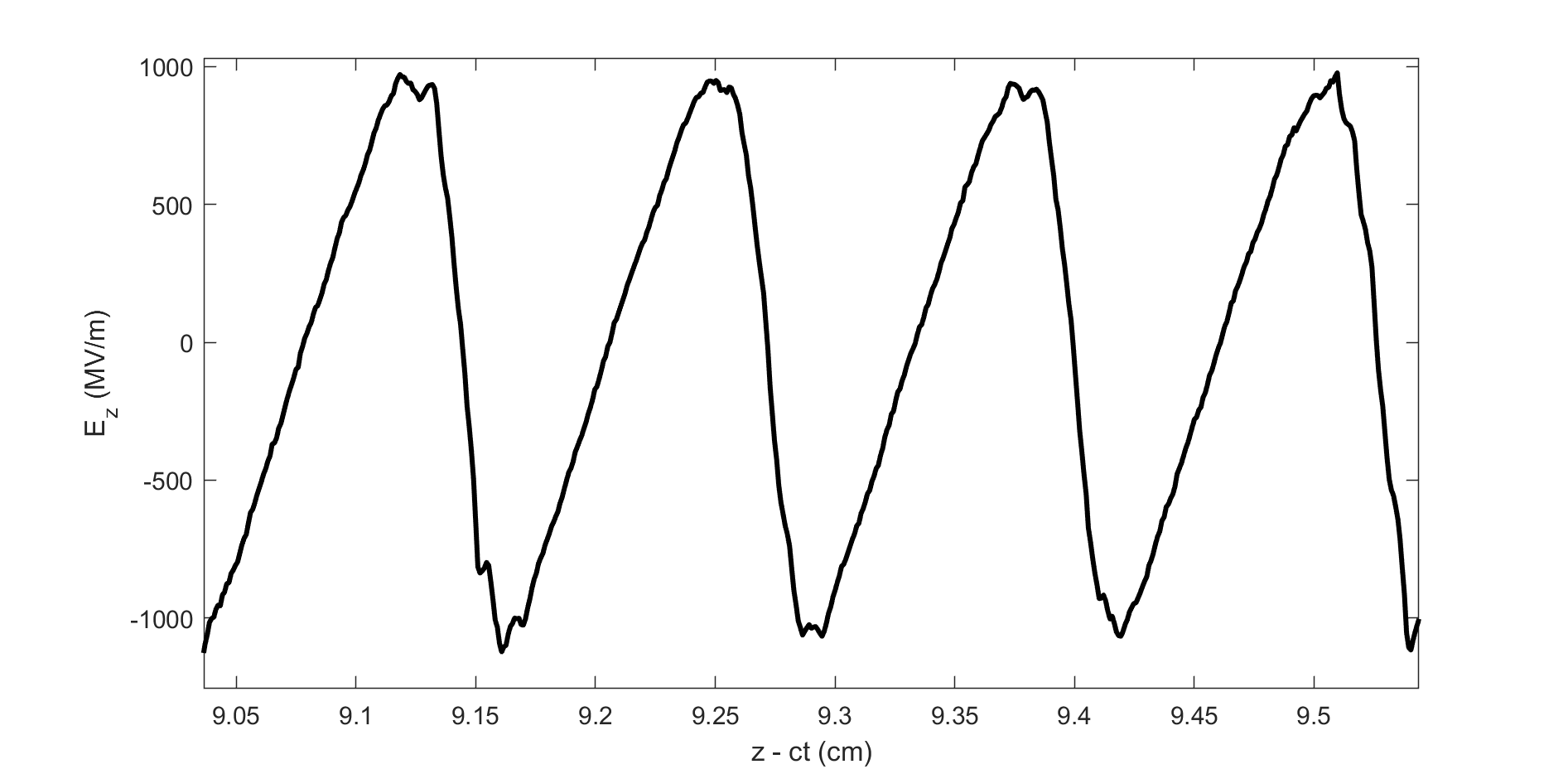}
\caption{Line-out of the longitudinal wakefields close to the axis at 200\,cm into the plasma. %
Wakefields at this point are non-sinusoidal and their amplitude reaches 0.37\,E$_{WB}$, i.e., are non-linear (seed bunch charge Q=3.69\,nC).} 
\label{fig:WR20} 
\end{figure}

In the high charge seed bunch case, seed wakefields are already large and quickly reach the non-linear regime, as seen in Fig.~\ref{fig:WR20} for the %transverse
longitudinal wakefields. % of the case with the highest seed bunch charge. % of 3.69\,nC. 
%One of the consequences of this is that the amplitude of the fields that defocus the proton  increase in size. %
A large fraction of the wakefields period ($>$50\%, the linear regime value) is defocusing for protons. %
This leads to a faster increase and steeper decrease of the maximum wakefields amplitude. %
For all cases, except the case with the maximum seed bunch charge, the system keeps evolving, expelling protons over the full plasma length, and producing wakefields of decreasing amplitude. 

In the case of the largest seed bunch charge (3.69\,nC), protons are expelled fast enough to only leave behind narrow micro-bunches that resonantly drive wakefields. %
This produces a system that drives wakefields with constant amplitude and that evolves very slowly ($z>$300\,cm). %
The proton bunch charge decreases continuously for the first 500\,cm of the plasma, thereafter it has a value %that oscillates between 9.6 and 9.8 \%
around 10\% of the initial bunch charge. %  

For completion, we also performed a simulation using the same conditions as with particle bunch seeding, but without seed bunch. % no bunch seed. %
The proton bunch is cut at three RMS lengths ahead of the bunch center, leading to low seed wakefields of 150\,kV/m amplitude (6$\times$10$^{-5}$\,E$_{WB}$) when compared to any of the other cases of seeding used previously. %
The SM evolves more slowly (orange lines on Fig.~\ref{fig:Ezmaxofz}). %
The longitudinal fields reach a peak amplitude at around 700\,cm, after which they start decreasing as in the other cases. %
%It is worth noting that this self-modulated bunch cannot be used for acceleration, because, in experiments, the fields would have a different phase from event to event~\cite{bib:fabian}. % cite Fabian's paper?  YES

Numerical studies are ongoing to have an in-depth understanding of the electron bunch seeding process and to determine how a plasma density step placed in the growth region of the SM process can maintain the longitudinal field at approximately its saturation values~\cite{bib:step} even with this seeding method. %

\subsection{Electron Bunch Seeding and Acceleration in a Single Plasma}

When conceiving of an accelerator based on wakefields driven by a self-modulated long bunch, one would naturally envisage separating the SM from the acceleration process. %
The self-modulation would be seeded with an electron bunch for the reasons outlined above. %
The electron bunch to be accelerated would be injected after the first short, self-modulation plasma, and on-axis into the long, acceleration plasma~\cite{bib:muggliRun2}. %
Since the accelerated bunch must create blow-out and load the wakefields~\cite{bib:veronica}, its wakefields amplitude must be similar to that of the self-modulated proton bunch. %
However, in practice, the injection section, between the two plasma sections, must be quite short, a difficult section to design, build and diagnose~\cite{bib:livio}. %

\begin{figure}[ht]
\centering
\includegraphics[scale=0.5]{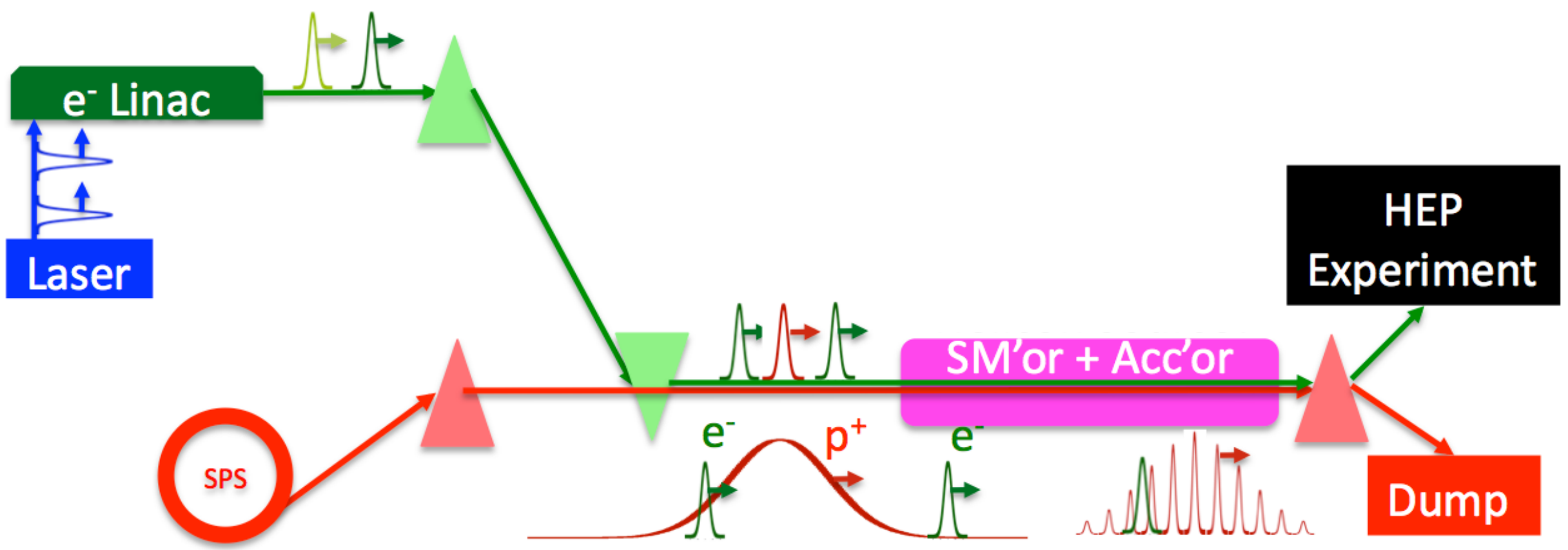}
\caption{Schematic set-up for injection of the three bunches, short electron seed, long proton and short accelerated electron bunches in a single plasma (not to scale). %
The seeding process leads to modulation of the entire proton bunch. %
The relative timing of the three bunches is for illustration only. %
Accelerated electrons go to a high-energy physics experiments, protons are dumped.} 
\label{fig:schematic2} 
\end{figure}

%scenario proposed here, the electron bunch parameters would be such that the bunch allows for blow-out to be reached and for loading of the wakefields driven by the self-modulated proton bunch~\cite{bib:veronica}. %
One could then envisage simplifying the plasma and injection scheme by injecting all three bunches right at the entrance of a single plasma, as shown on Fig.~\ref{fig:schematic2}. %
Because the accelerated bunch drives wakefields with amplitudes similar to those of the self-modulated proton bunch~\cite{bib:veronica}, it could remain weakly affected by the proton bunch wakefields both in the (probable) density ramp at the plasma entrance and by the growing wakefields. %
Moreover, its relative timing with respect to the seed electron bunch and proton bunch could be chosen to minimize these effects (if any). %

The seed electron bunch would lose most of its energy and thus drive wakefields that would be overwhelmed by those of the self-modulated proton bunch. % (saturated to seed wakefields amplitude ratio $\gg$10:1). %
It would therefore disappear naturally after some distance, as low energy electrons, and not affect the wakefields or the proton bunch after that. %

We are currently investigating all these possibilities in numerical simulations. %
We are looking for optimum input parameters and ultimate output parameters in terms of charge, relative energy spread, and emittance of the accelerated electron bunch. %
We will also investigate tolerances in terms of relative transverse alignment of the three bunches so as not to seed the hose instability of the proton or accelerated electron bunch, and reach good accelerated bunch parameters. %

This scheme would allow for the use of a single, pre-ionized plasma that could be made as long as necessary for high-energy physics applications, for example by using a single, very long helicon plasma source~\cite{bib:helicon}. %

\section{Summary}

We described seeding methods for the self-modulation of a long charged particle bunch in plasma, as they apply in particular to the AWAKE experiment. %
We briefly compared relativistic ionization front seeding, emulated in numerical simulations by a cut bunch, with electron bunch seeding in simple cases. %
We showed that the evolution of the wakefields as well as their parameters can be similar in both seeding schemes. %
We showed that the value and position of the peak of the maximum longitudinal electric field can be affected by varying the seed bunch charge and thus the seed wakefields amplitude. %
Electron bunch seeding could allow for simplification of the electron injection scheme and allow for a single plasma source to be used. %
The work presented here is the seed for more studies to develop a simple concept for accelerating electrons to high-energy for high-energy physics applications. %

\end{document}